# Influence of Preservation Temperature on the Measured Mechanical Properties of Brain Tissue

Badar Rashid[a], Michel Destrade[b,a], Michael Gilchrist[a*]

[a]School of Mechanical and Materials Engineering, University College Dublin, Belfield, Dublin 4, Ireland

[b]School of Mathematics, Statistics and Applied Mathematics, National University of Ireland Galway, Galway, Ireland

*Corresponding Author

Tel: + 353 1 716 1884/1991, + 353 91 49 2344  Fax: + 353 1 283 0534

Email: Badar.Rashid@ucdconnect.ie (B. Rashid), michael.gilchrist@ucd.ie (M.D. Gilchrist), michel.destrade@nuigalway.ie (M. Destrade)

**Abstract** The large variability in experimentally measured mechanical properties of brain tissue is due to many factors including *heterogeneity, anisotropy, age dependence* and *post-mortem time*. Moreover, differences in test protocols also influence these measured properties. This paper shows that the temperature at which porcine brain tissue is stored or preserved prior to testing has a significant effect on the mechanical properties of brain tissue, even when tests are conducted at the same temperatures. Three groups of brain tissue were stored separately for at least one hour at three different preservation temperatures, i.e., ice cold, room temperature (22°C) and body temperature (37°C), prior to them all being tested at room temperature (~22°C). Significant differences in the corresponding initial elastic shear modulus $\mu$ (Pa) (at various amounts of shear, $K$, i.e., 0–0.2) were observed. The initial elastic moduli were 1043±271 Pa, 714±210 Pa and 497±156 Pa (mean±SD) at preservation temperatures of ice cold, 22°C and 37°C, respectively. Based on this investigation, it is strongly recommended that brain tissue samples must be preserved at an ice-cold temperature prior to testing in order to minimize the difference between the measured *in vitro* test results and the *in vivo* properties. A by-product of the study is that simple shear tests allow for large, almost perfectly homogeneous deformation of brain matter.

# 1    Introduction

Extensive research has been carried out over the past five decades to characterize the mechanical properties of brain tissue in order to establish realistic constitutive relationships over a wide range of loading conditions. In particular, there is a pressing need to characterize brain tissue properties over the expected loading rate associated with traumatic brain injuries (TBIs). However, the reliable determination of brain tissue properties is a formidable challenge, as it depends heavily on various experimental parameters.

A limited number of studies has investigated the effects of variable temperatures (Brands et al., 2000; Peters et al., 1997; Shen et al., 2006). Hrapko et al. (2008) stored brain samples in phosphate buffered saline (PBS) in a box filled with ice during transportation and maintained at ~4°C before testing. Tests were conducted at room temperature (23°C) and at body temperature (37°C). The measured results were clearly temperature dependent and the dynamic modulus $G^*$ was 60% higher at 23°C than at 37°C. This clearly indicates that testing brain tissue at higher temperature accelerates degradation of the mechanical integrity of tissue, thus further deviating from *in vivo* test conditions. However, in a different study by Zhang et al. (2011), brain samples were preserved in ice cold (group A, 10 samples) and in 37°C (group B, 9 samples) saline solutions. All samples were warmed to a temperature of 37°C in a saline bath prior to testing. The stress response from brain samples preserved at 37°C was 2.4 times stiffer at 70% strain, than when preserved at the ice-cold temperature. These findings directly contradict the study by Hrapko et al. (2008), thus leading to inconclusive results and raising important questions: Do higher temperatures lead to a stiffer or a softer response? At which temperature should brain samples be stored prior to testing? In the literature, protocols vary greatly: for instance, Pervin and Chen (2009; 2011) stored tissues at 37°C, whereas Miller and Chinzei (1997; 2002), Tamura et al., (2008; 2007) and Rashid et al., (2012b,c,d) stored brain tissues at ice cold /4–5°C before the tests. The reliability of experimental data obtained from the tissue preserved at higher temperature (37°C) is questionable based on the contradictory findings of existing studies (Zhang et al. (2011); Hrapko et al. (2008)). It is, therefore, crucial to clearly understand the behavior of tissue under different preservation temperature conditions, with a view to achieving reliable material parameters.

With this aim in mind, simple shear tests were performed on brain tissue at a strain rate of 30/s (i.e., 3000%/s, *not* 30%/s) and up to 62% engineering shear strain (amount of shear, $K = 1$, where K is the ratio of horizontal displacement of the top of a specimen of brain tissue to its thickness, as indicated in Section 2.1 below) under different temperature conditions. It is important to realize for simple shear that material is being deformed in various directions at different rates, and so the strain rate around a point within a material cannot be expressed by a single number. We took the strain rate tensor to be the symmetric part of the velocity gradient and not the time derivative of the strain tensor. Three groups of brain tissue were stored separately for at least an hour at three different preservation temperatures: ice cold, room temperature (22°C) and body temperature (37°C), whereas experimentation was performed at an approximately constant room temperature (~22°C). The simple shear test protocol is adopted here because of its high reliability due to a global homogeneous deformation field of brain tissue as compared to compression and tension test protocols, which lead to inhomogeneous deformation fields (Ogden, 1997; Rashid et al., 2012a).

## 2 Materials and Methods

### 2.1 Simple Shear Experimental Setup

A High Rate Shear Device (HRSD) as described in Fig. 1 (a) and (b) was used to perform simple shear tests at a dynamic strain rate of 30/s (i.e., 3000%/s). The development and major components of the HRSD have been discussed elsewhere (Rashid, 2012). During tests, the top platen remained stationary while the lower platen moved horizontally to produce the required simple shear deformation in the specimen, as shown in Fig. 1. Force (N) and displacement (mm) signals were captured simultaneously through the data acquisition system at a sampling rate of 10 kHz. The amount of shear is $K = d/y$, where $d$ is the horizontal displacement of the lower platen (maximum displacement is 4.0 mm) and $y$ is the thickness of the specimen (4.0 mm for all tests). Therefore, $K = 1$ is the maximum amount of shear or shear strain. The intended velocity of the electronic actuator was 120 mm/s. Hence, the lower platen travelled 4.0 mm horizontally in 1/30 s, to achieve a maximum amount of shear K = 1 for a 4.0 mm thick specimen. This gave a shear component for the strain rate tensor (symmetric part of the velocity gradient) of 30 s$^{-1}$, which we note as 30/s; this rate is typical of TBIs. However, the actual loading velocity was slightly higher (130 mm/s) in order to overcome the frictional effects and opposing spring force acting against the *striker*, which were adjusted during the calibration process.

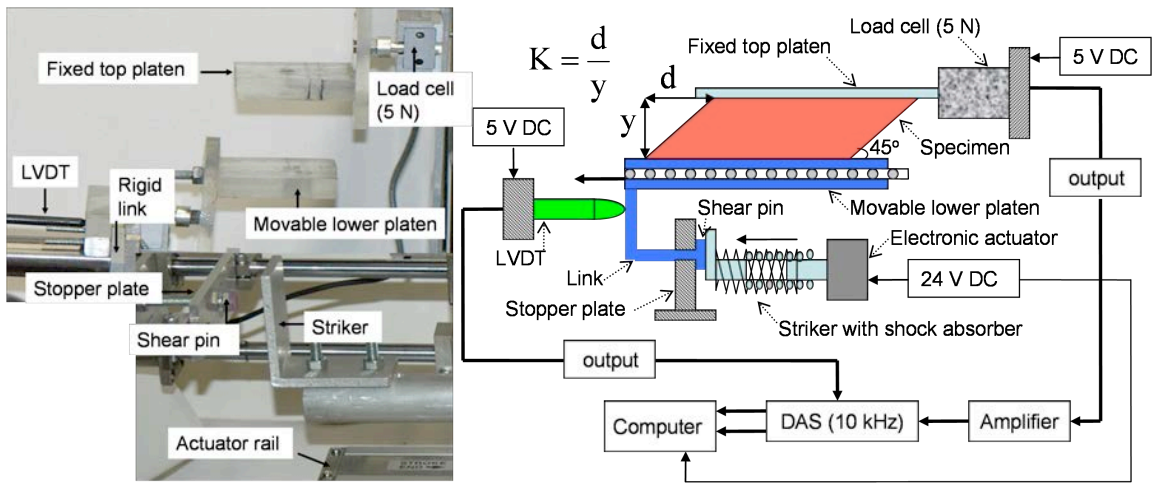

Fig. 1 – (a) Major components of high rate shear device (HRSD),
(b) Schematic diagram of complete test setup, with $K = 1$ for maximum amount of shear.

### 2.2 Specimen Preparation Procedure

Nine fresh porcine brains from approximately six-month old pigs were collected from a local slaughterhouse and tested within 5 h postmortem. The brains were divided into three groups which were preserved for one hour in a physiological saline solution at three different temperatures (3 brains each in ice cold, 22°C and 37°C) during transportation. All samples were prepared and tested at a nominal room temperature of 22°C. Square specimens as shown in Fig 2, composed of mixed white and gray matter were prepared using a square steel cutter after removing the dura and arachnoid from the cerebral hemispheres. Two specimens were extracted from each cerebral hemisphere from the medial to lateral direction. The thickness, width and length of specimens before testing were 4.0±0.2 mm, 19.0±0.1 mm and 19.0±0.1 mm (mean±SD), respectively. 36 specimens were prepared from the 9 brains (4 specimens from each brain). The time elapsed between harvesting of the first and last specimens from each brain was approximately 18 minutes. Physiological saline solution was applied to the specimens frequently during cutting and before the tests in order to prevent dehydration.

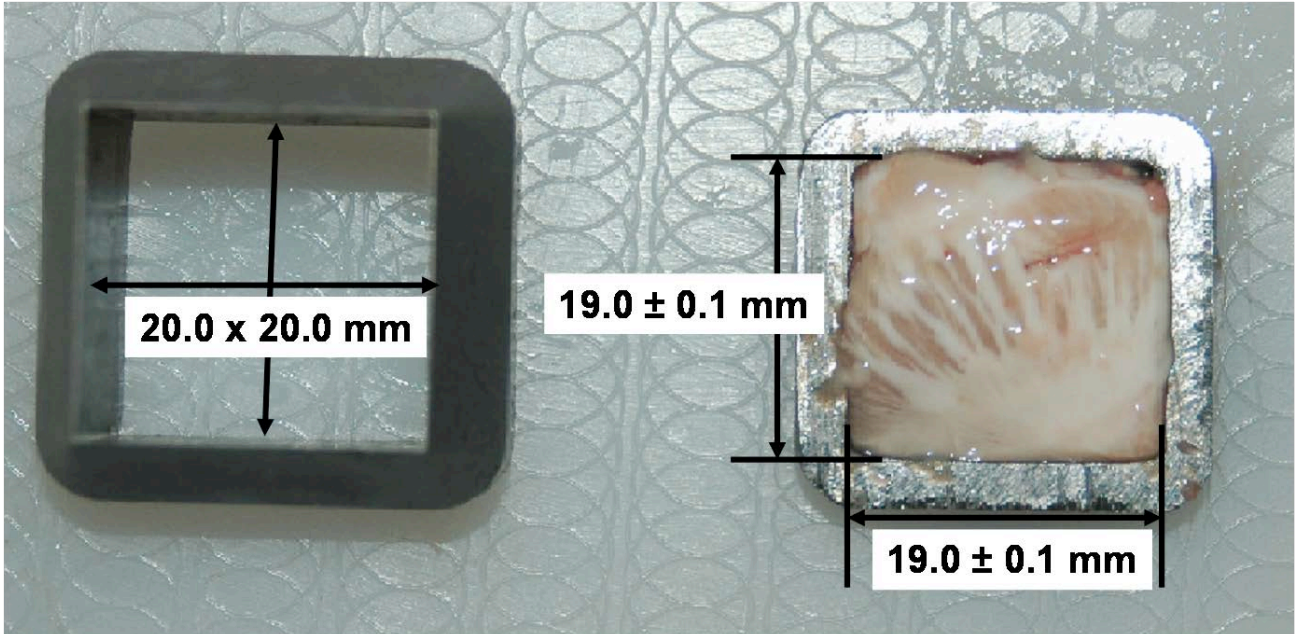

Fig. 2 – Square brain specimen (19.0±0.1 x 19.0±0.1 mm) and 4.0±0.1 mm thick excised from medial to lateral direction.

## 2.3  Specimen Attachment Procedure

The surfaces of the platens were first covered with a masking tape substrate to which a thin layer of surgical glue (Cyanoacrylate, Low-viscosity Z105880–1EA, Sigma-Aldrich) was applied. The prepared specimens of brain tissue were then placed on the lower platen. The top platen was attached to the 5 N load cell, and was then lowered slowly so as to just touch the top surface of the specimen. One minute settling time was sufficient to ensure proper adhesion of the specimen to the platens. Before mounting the brain specimens for simple shear tests, calibration of the HRSD was performed to ensure uniform velocity at a strain rate of 30/s. The top platen was fixed, while the lower platen was restricted to move horizontally, so that no normal expansion or shortening of the specimen's thickness was allowed to take place. In effect, the constraints on the platens imposed normal stresses (along with the shear stress) in order to counteract the Poynting effect of nonlinear elasticity.

# 3     Results

## 3.1    Simple Shear Experiments

Ten tests were performed for each preservation temperature condition (ice-cold, room temperature: 22°C and body temperature: 37°C) at a strain rate of 30/s, as shown in Fig. 3. The force (N), sensed by the load cell attached to the top platen, was divided by the surface area in the reference configuration to determine the shear stress (Pa). Similarly, the displacement was divided by the original thickness of the specimen to determine the amount of shear $K$. During simple shear tests, the achieved strain rate was 30±1.65 /s, (mean±SD) against the required loading velocity of 120 mm/s.

The maximum shear component of the Cauchy stress (at maximum amount of shear, $K = 1$) at preservation temperatures of ice cold, 22 and 37°C was 1545±383 Pa, 1019±295 Pa and 699±195 Pa (mean±SD), respectively, as shown in Fig. 3.

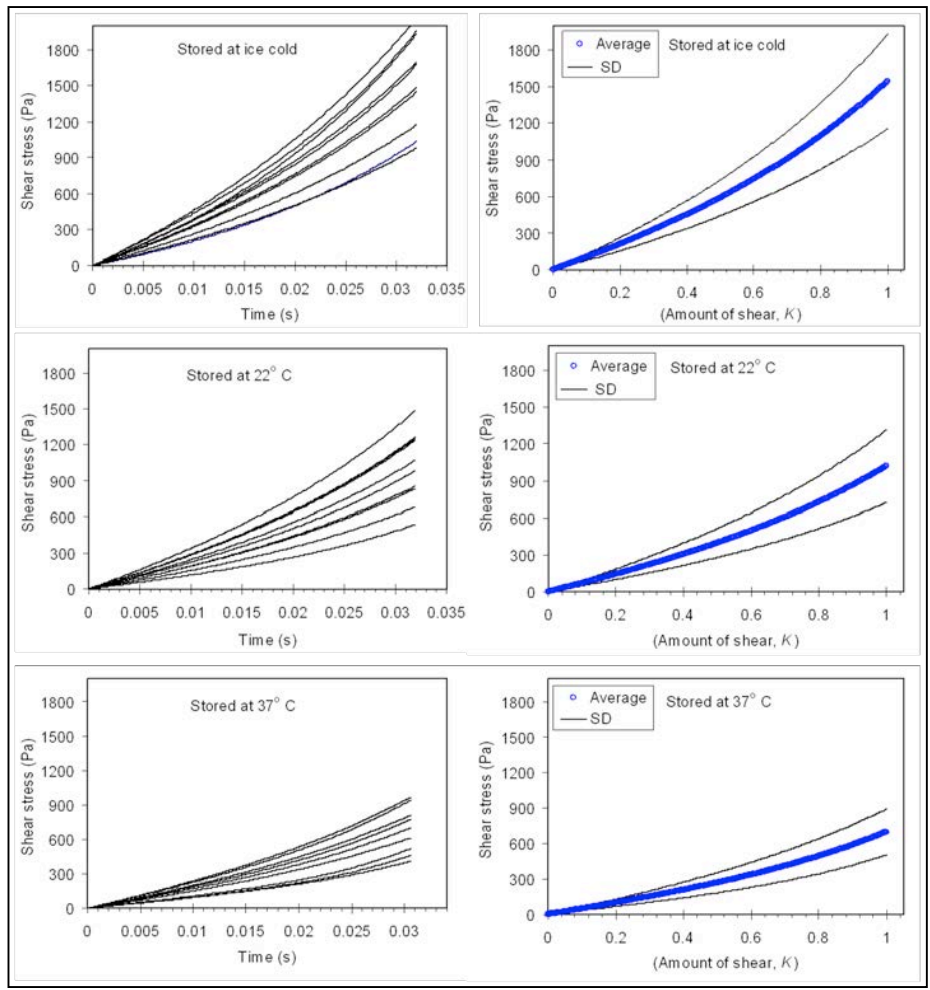

Fig. 3 – Variation in Cauchy shear stress magnitudes of brain tissue at variable preservation temperature conditions (ice cold, 22°C and 37°C), tested at a strain rate of 30/s.

## 3.2 Hyperelastic Material Parameters

Now it is useful to estimate the nonlinear material parameters based on simple shear data at variable temperature conditions. In the Rectangular Cartesian coordinate system aligned with the edges of the undeformed specimen, the simple shear deformation can be written as

$$x_1 = X_1 + KX_2, \quad x_2 = X_2, \quad x_3 = X_3 \tag{1}$$

where $K$ is the amount of shear, $\underline{x}$ is the current coordinate and $\underline{X}$ the reference coordinate. Using Eq. (1), the deformation gradient tensor $\underline{\mathbf{F}}$ and the right Cauchy-Green deformation tensor $\underline{\mathbf{C}} = \underline{\mathbf{F}}^T\underline{\mathbf{F}}$ are

$$\underline{\mathbf{F}} = \begin{bmatrix} 1 & K & 0 \\ 0 & 1 & 0 \\ 0 & 0 & 1 \end{bmatrix}, \quad \underline{\mathbf{C}} = \underline{\mathbf{F}}^T\underline{\mathbf{F}} = \begin{bmatrix} 1 & K & 0 \\ K & 1+K^2 & 0 \\ 0 & 0 & 1 \end{bmatrix} \tag{2}$$

In general, an isotropic hyperelastic incompressible material is characterized by a strain-energy density function $W$ which is a function of two principal strain invariants only: $W = W(I_1, I_2)$, where $I_1$ and $I_2$ as defined as (Ogden, 1997)

$$I_1 = \mathrm{tr}(\mathbf{C}), \quad I_2 = \frac{1}{2}[I_1^2 - \mathrm{tr}(\mathbf{C}^2)], \tag{3}$$

But in the present case of simple shear deformation,

$$I_1 = I_2 = 3 + K^2 \tag{4}$$

so that
$$W = W(3+K^2, 3+K^2) \equiv \hat{W}(K) \text{ say.} \tag{5}$$

The shear component of the Cauchy stress tensor is (Ogden, 1997):

$$\sigma_{12} = 2K\left(\frac{\partial W}{\partial I_1} + \frac{\partial W}{\partial I_2}\right) = \hat{W}'(K) \tag{6}$$

The Cauchy shear stress component $\sigma_{12}$ was evaluated as $\sigma_{12} = F/A$, where $F$ is the shear force, and $A$ is the area of a cross section of the specimen, which remains unchanged in simple shear (so that $\sigma_{12} = S_{12}$, the nominal shear stress component). The experimentally measured shear stress component was then compared to the predictions of the hyperelastic model from the relation $\sigma_{12} = \hat{W}'(K)$ (Ogden, 1997), and the material parameters were adjusted to give good curve fitting. The fitting was performed using the *lsqcurvefit.m* function in MATLAB. The shear response curves in Fig. 3 clearly showed a non-linear relationship, which ruled out the Mooney-Rivlin and neo-Hookean models. We chose the one-term Ogden hyperelastic model because it gave excellent fitting with only two fitting parameters. It is given by

$$W = \frac{2\mu}{\alpha^2}\left(\lambda_1^\alpha + \lambda_2^\alpha + \lambda_3^\alpha - 3\right) \qquad (8)$$

where the $\lambda_i$ are the principal stretch ratios (the square roots of the eigenvalues of $\underline{C}$), $\mu > 0$ is the infinitesimal shear modulus, and $\alpha$ is a stiffening parameter. In simple shear:

$$\lambda_1 = \frac{K}{2} + \sqrt{1 + \frac{K^2}{4}},\ \lambda_2 = \lambda_1^{-1} = -\frac{K}{2} + \sqrt{1 + \frac{K^2}{4}},\ \lambda_3 = 1, \qquad (9)$$

and the Cauchy shear stress component $\sigma_{12}$ is thus

$$\sigma_{12} = \hat{W}'(K) = \frac{\mu}{\alpha}\frac{1}{\sqrt{1+\frac{K^2}{4}}}\left[\left(\frac{K}{2}+\sqrt{1+\frac{K^2}{4}}\right)^\alpha - \left(-\frac{K}{2}+\sqrt{1+\frac{K^2}{4}}\right)^\alpha\right]. \qquad (10)$$

When $\alpha = 2$, it reduces to a linear relationship, $\sigma_{12} = \mu K$, because the $W$ function is then that of the neo-Hookean model. When $\alpha < 2$, the material softens in shear. Here, it clearly stiffens in shear, and we thus expect that $\alpha > 2$. The material parameters $\mu$ and $\alpha$ derived after fitting Eq. (10) to average shear stress – engineering shear strain profiles (Fig. 3) are summarized in Table 2. Sometimes the waviness in experimental data affected the values of the stiffening parameter $\alpha$ and we then used the data smoothing capabilities of Matlab to reduce the influence of data noise. Note that the largest stretch was attained when $K = 1$ in our experiments, giving $\lambda_1 = 1.618$, according to Eq. (9), i.e., a strain of 62%. We observed a significant decrease in initial shear modulus $\mu$ with the increase in temperature as shown in Table 1, while the (nonlinear) stiffening parameter $\alpha$ remained almost constant.

### 3.3 Finite Element Simulations

In order to check whether the assumption of homogeneous simple shear was reasonable for our dimensions and protocols, we prepared a brain tissue specimen geometry in ABAQUS 6.9 to mimic experimental conditions (discussed in Section 2.2). We used 2166 x C3D8R elements, default hourglass control, mass density of 1040 kg/m$^3$ and material parameters listed in Table 1 for the numerical simulations. The top surface of the specimen was constrained in all directions whereas the lower surface was allowed to move only in the lateral direction ($x_1$- axis) in order to reach the maximum amount of shear, $K = 1$. Visual inspection, coupled to a one-way ANOVA test, revealed excellent agreement between the average experimental and numerical shear stresses (p = 0.9215 at ice cold, p = 0.9333 at 22°C and p = 0.8489 at 37°C), as shown in Fig. 4 (a). A significant difference (p = 0.0004) existed between the shear stresses at ice cold and 37°C, as clearly depicted in Fig. 4 (a) and (b). There was a 34% decrease in $\mu$ from ice cold to 22°C and a 31.4% decrease from 22°C to 37°C, which clearly indicates the stiffening response of brain tissue with lower preservation temperatures. An almost homogeneous deformation was achieved throughout the sample during experimentation and during numerical simulations according to Figs. 4 (c) and (d), which gives credence to the reliability of the simple shear test protocol.

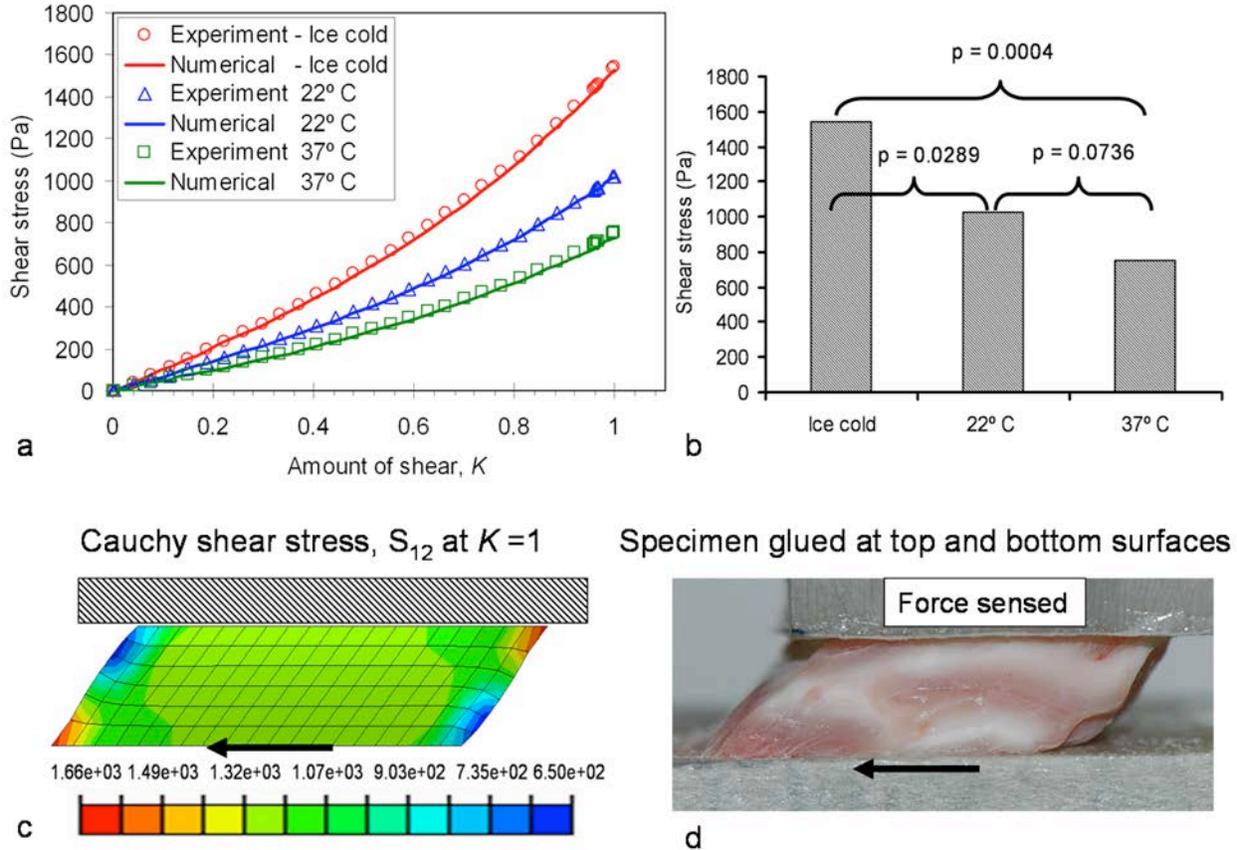

Fig. 4 – (a) Excellent agreement between the experimental and numerical shear stress values at different temperatures (b) shows results of one-way AVOVA analysis based on average experimental shear stress values (c) homogeneous deformation observed in numerical simulations (under ice cold condition) (d) homogeneous deformation of brain tissue during simple shear experiments.

## 4   Discussion

The material parameters obtained from fitting the shear response curve to the predictions of a one-term Ogden strain energy function showed a significant increase in initial shear modulus, $\mu$, with lower preservation temperatures: $\mu$ = 500.7 Pa for 37°C, $\mu$ = 715.3 Pa for 22°C, and $\mu$ = 1050 Pa for ice-cold. The stresses for the ice-cold preservation temperature are 1.5 times higher than for body temperature (37°C), while performing tests at the same room temperature (22°C) and at the maximum amount of shear ($K$=1), as clearly shown in Fig 4 (a).

Miller and Chinzei (1997) performed *in vitro* unconfined compression tests on porcine brain tissue having 5°C as the preservation temperature. However, the forces measured during *in vivo* indentation tests (Miller et al., 2000) were 31% higher than during *in vitro* tests (Miller and Chinzei, 1997). Presumably, the existing difference (31%) between the *in vivo* and *in vitro* results would increase further if brain tissue was preserved at higher temperatures (37°C), because of the decrease in tissue stiffness clearly observed in our experiments (see Fig. 4 (a)). Conversely, if brain tissue was preserved at a lower (ice-cold) temperature, then its stiffness would increase and the gap of 31% would be reduced. Therefore, an ice cold /(4–5°C) preservation temperature is necessary to minimize the difference between *in vitro* and *in vivo* results and to partially compensate for the loss of stiffness due to the release of residual stress when extracting samples.

Garo et al. (2007) found no significant changes in the mechanical properties (p = 0.95) of the brain tissue for samples tested between 2 and 6 h *post-mortem*. Similarly, McElhaney et al. (1973) reported no significant changes up to 15 h *post-mortem* and Nicolle et al. (2004) found only a 6% increase in the linear viscoelastic response for samples tested at 24 and 48 h of *post-mortem*. Darvish and Crandall (2001) found no correlation between time and variation in mechanical properties for the tests conducted between 3–16 days later. Only the study conducted by Metz et al. (1970) reported a 30–70% decrease in the tissue response from live to 3–4 hour *post-mortem* time. In the present study, all tests were completed within 5 h of *post-mortem* in order to minimize the possibility of variations in experimental data potentially linked to the post-mortem time interval.

The use of a thin layer of surgical glue (approximately one drop on each platen) proved reliable for the attachment of brain tissue and did not alter the stiffness of the tissue. This factor was further investigated by performing a separate set of simple shear experiments with variable specimen thicknesses (~2.0, 3.0, 4.0, 5.0, 6.0 mm) at the same strain rate (30/s). No difference in results was observed, thus proving the reliability of this test protocol (as discussed in Section 2.3). Moreover, finite element simulations were also performed using the one-term Ogden parameters ($\mu$ = 1050 Pa, $\alpha$ = 4.1 for ice cold conditions), as shown in Fig. 5. Excellent agreement between the shear stress profiles at variable specimen thickness was achieved (p = 0.9978 based on one way ANOVA).

A limitation of this study is the estimation of *global* material parameters from the strain energy functions, based on *average* mechanical properties (mixed white and gray matter) of the brain tissue. However, our results are still useful in modelling the approximate behaviour of brain tissue, in line with the procedure followed by Miller and Chinzei (1997; 2002).

In order to fully characterize the behavior of brain tissue, viscoelastic tests (stress relaxation tests) are usually performed to obtain one set of parameters; however, the deterioration in tissue properties due to higher preservation temperature affects both the hyperelastic and the viscoelastic parameters. Therefore in this study, only the hyperelastic behavior of brain tissue was considered so as to investigate the effects of preservation temperature in isolation.

The following conclusions can be made from this study:

1. Measured brain tissue properties are significantly influenced by the preservation temperatures (one-term Ogden initial shear modulus, $\mu$ = 500.7 Pa at 37°C, $\mu$ = 715.3 Pa at 22°C and $\mu$ = 1050 Pa for ice cold preservation temperature).

2. Brain tissue must be preserved at an ice cold temperature prior to testing in order to minimize the difference between the measured *in vitro* and actual *in vivo* properties.

3. The simple shear test is most suitable for the reliable collection of results, based on an almost perfectly homogeneous deformation field, even at large strains (up to 60%).

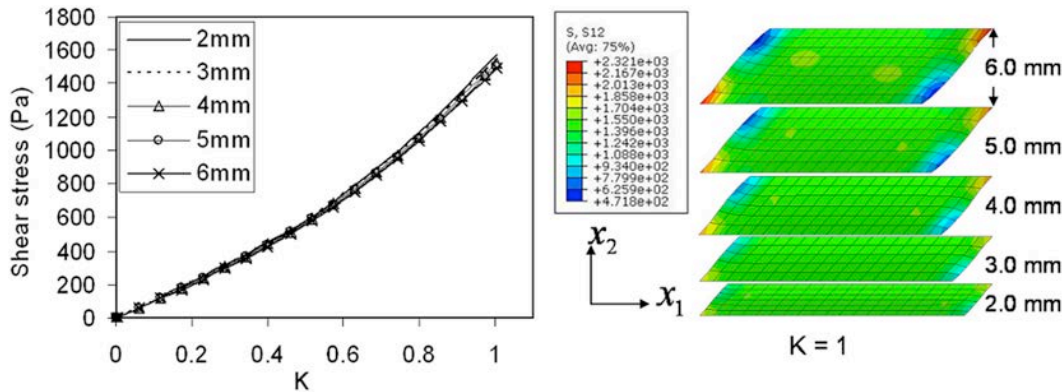

Fig. 5 – Consistency in shear stress profiles using Ogden material parameters obtained at a strain rate of 30/s.